\documentstyle[epsfig]{aipproc}
\newcommand{\replabel}[1]{\hbox{\raise 12pt
  \hbox to\hsize{\hfill\normalsize\rm #1}}}

\begin{document}
\title{\replabel{MZ-TH/99-18, CLNS/99-1619, hep-ph/9905373, May 1999}Local
  and Global Duality and the Determination of $\alpha(M_Z)$}
\author{Stefan Groote}
\address{Institut f\"ur Physik, Johannes-Gutenberg-Universit\"at,
Staudinger Weg 7, 55099 Mainz, Germany\\
Floyd R.~Newman Laboratory of Nuclear Studies,
Cornell University, Ithaca, NY 14853, USA}
\maketitle
\kern-24pt
\begin{abstract}\noindent
This talk\footnote{Talk given at the conference ``High Energy Physics at the
Millennium (MRST'99)'', Ottawa, Canada, May 10--12, 1999}
presents work concepts and results for the determination of the
fine structure constant $\alpha$ at the $Z_0$ mass resonance. The problem
consisting of the break-down of global duality for singular integral weights
is circumvented by using a polynomial fit which mimics this weight function.
This method is conservative in the sense that it is mostly independent of
special assumptions. In this context the difference between local and global
duality is explained.
\end{abstract}

\section*{Introduction}
There is a great deal of interest in the accurate determination of the
running fine structure constant $\alpha$ at the scale of the $Z_0$
mass~\cite{Groote:ref1,Groote:ref2,Groote:ref3,Groote:ref4}. The
value of $\alpha(M_Z)$ is of paramount importance for all precision tests of
the Standard Model. Furthermore, an accurate knowledge of $\alpha(M_Z)$ is
instrumental in narrowing down the the mass window for the last missing
particle of the Standard Model, the Higgs particle.

The main source of uncertainty in the determination of $\alpha(M_Z)$ is the
hadronic contribution to the $e^+e^-$ annihilations needed for this
evaluation. The necessary dispersion integral that enters this calculation
has in the past been evaluated by using experimental $e^+e^-$ annihilation
data. Discrepancies in the experimental data between different experiments
suggest large systematic uncertainties in each of the experiments. In order
to reduce the influence of the systematic uncertainties on the determination
of $\alpha(M_Z)$ one may attempt to add some theoretical input to the
evaluation of the hadronic contribution to $\alpha(M_Z)$.

M.~Davier and A.~H\"ocker~\cite{Groote:ref3} use QCD perturbation theory
in form of local duality (explained later on) in the region above
$s=(1.8{\rm\,GeV})^2$ for the light flavours, while J.H.~K\"uhn and
M.~Steinhauser~\cite{Groote:ref5} use perturbative results for energy
regions outside the charm and bottom threshold regions. Our
approach~\cite{Groote:ref6} is quite different. We attempt to minimize the
influence of data in the dispersion integral over the whole energy region
including the threshold regions.

\subsection*{Local and global duality}
For the $e^+e^-$ annihilation process there is a main connection between
the spectral density $\rho(s)$ and the two-point correlator $\Pi(q^2)$
given in kind of the dispersion relation
\begin{equation}
\Pi(q^2)=\int_{s_0}^\infty\frac{\rho(s)ds}{s+q^2}
\end{equation}
which implies the reverse relation
\begin{equation}
\rho(s)=\frac1{2\pi i}{\sl Disc\,}\Pi(s)
\end{equation}
where the discontinuity is given by
\begin{equation}
{\sl Disc\,}\Pi(q^2)=\Pi(q^2e^{-i\pi})-\Pi(q^2e^{i\pi})
\end{equation}
and $s_0=4m_\pi^2$ is the production threshold of the light flavours.
These relations can be a chain between the theory, i.e.\ the two-point
correlator function within perturbative QCD on the one hand and the
experiment, i.e.\ the spectral density or, equivalently, the total cross
section on the other hand. But there is one obstacle in using these
relations: As depending on methods of functional analysis, the inverse of
the dispersion relation is only valid if there are no poles encircled by
the path in the complex plane, a condition which is necessary to obtain
this relation. These poles can have their origin from weight functions in
combination with the spectral density. This means that if there is such a
weight function included in the integration of the spectral density, the
inverse relation shown above is only valid {\em locally\/} and not {\em
globally\/}. We call this {\em local\/} resp.\ {\em global duality}.

\subsection*{The experiment side}
The hadronic contribution to $\alpha(M_Z)$ which we are concentrating on
is given by the integral~\cite{Groote:ref1}
\begin{equation}
\Delta\alpha_{\rm had}(M_Z)=\frac\alpha{3\pi}{\sl Re\,}\int_{s_0}^\infty
  R(s)H(s)ds
\end{equation}
where $R(s)$ is the total $e^+e^-$ hadronic cross section and $H(s)$ is the
weight function
\begin{equation}
H(s)=\frac{M_Z^2}{s(M_Z^2-s)}.
\end{equation}
The hadronic cross section is related to the spectral density by
\begin{equation}
R(s)=12\pi^2\rho(s).
\end{equation}
But we see: the weight function $H(s)$ is indeed singular at the points
$s=0$ and $s=M_Z^2$ on the real axis. So global duality is not valid any
more.

\subsection*{The theory side}
The two-point correlator is given by
\begin{equation}
i\int\langle 0|j_\alpha^{\rm em}(x)j_\beta^{\rm em}(0)|0\rangle e^{iqx}d^4x
  =(-g_{\alpha\beta}q^2+q_\alpha q_\beta)\Pi(q^2)
\end{equation}
where we only included the isospin contribution $I=1$, in contrast to
corresponding considerations for the $\tau$ decay. The scalar correlator
function $\Pi(q^2)$ consists of perturbative and non-perturbative
contributions which we include to the extend we need them to keep the
accuracy. For the perturbative contribution to the correlator we use a
result given in ref.~\cite{Groote:ref7}. I only write down the first few
terms,
\begin{equation}
\Pi^{\rm P}(q^2)=\frac3{16\pi^2}\sum_{i=1}^{n_f}Q_i^2\Bigg[
  \frac{20}9+\frac43L+C_F\left(\frac{55}{12}-4\zeta(3)+L\right)
  \frac{\alpha_s}\pi+O(\alpha_s^2,m_q^2/q^2)\Bigg]
\end{equation}
with $L=\ln(\mu^2/q^2)$ while in ref.~\cite{Groote:ref7} the expression is
given up to $O(\alpha_s^2,m_q^{12}/q^{12})$. The number of active flavours
is denoted by $n_f$. For the zeroth order term in the $m_q^2/q^2$ expansion
we have added higher order terms in $\alpha_s$,
\begin{equation}
  \frac3{16\pi^2}\sum_{i=1}^{n_f}Q_i^2\Bigg[\left(c_3+3k_2L
  +\frac12(k_0\beta_1+2k_1\beta_0)L^2\right)\left(\frac{\alpha_s}\pi\right)^3
  +O(\alpha_s^4)\Bigg]
\end{equation}
with $k_0=1$, $k_1=1.63982$ and $k_2=6.37101$. We have denoted the yet unknown
constant term in the four-loop contribution by $c_3$. Remark, however, that
the constant non-logarithmic terms will not contribute to our calculations.
The non-perturbative contributions are given in ref.~\cite{Groote:ref8},
\begin{eqnarray}
\Pi^{\rm NP}(q^2)&=&\frac1{18q^4}\left(1+\frac{7\alpha_s}{6\pi}\right)
  \langle\frac{\alpha_s}\pi G^2\rangle\nonumber\\&&
  +\frac8{9q^4}\left(1+\frac{\alpha_s}{4\pi}C_F+\ldots\ \right)
  \langle m_u\bar uu\rangle
  +\frac2{9q^4}\left(1+\frac{\alpha_s}{4\pi}C_F+\ldots\ \right)
  \langle m_d\bar dd\rangle\nonumber\\&&
  +\frac2{9q^4}\left(1+\frac{\alpha_s}{4\pi}C_F
  +(5.8+0.92L)\frac{\alpha_s^2}{\pi^2}\right)
  \langle m_s\bar ss\rangle\nonumber\\&&
  +\frac{\alpha_s^2}{9\pi^2q^4}(0.6+0.333L)
  \langle m_u\bar uu+m_d\bar dd\rangle\nonumber\\&&
  -\frac{C_Am_s^4}{36\pi^2q^4}\left(1+2L+(0.7+7.333L+4L^2)\frac{\alpha_s}\pi
  \right)\\&&
  -\frac{448\pi}{243q^6}\alpha_s|\langle\bar qq\rangle|^2+O(q^{-8})\nonumber
\end{eqnarray}
where we have included the $m_s^4/q^4$-contribution arising from the unit 
operator. In this expression we used the $SU(3)$ colour factors $C_F=4/3$, 
$C_A=3$, and $T_F=1/2$. For the coupling constant $\alpha_s$ as well as for
the running quark mass we use four-loop expression given in
refs.~\cite{Groote:ref9,Groote:refa,Groote:refb} even though in both cases
the three-loop accuracy would already have been sufficient for the present
application.

\section*{Introducing our method}
Our method is based on the fact that we can use global duality when the
weight function is non-singular. This is the case for a polynomial function.
So we mimic the weight function by a polynomial function obeying different
conditions which we will explain later. By adding and subtracting this
polynomial function $P_N(s)$ of given order $N$ to the weight function
$H(s)$, we obtain without any restrictions
\begin{equation}\label{Groote:splitting}
\int_{s_a}^{s_b}\rho(s)H(s)ds=\int_{s_a}^{s_b}\rho(s)
  \left(H(s)-P_N(s)\right)ds+\int_{s_a}^{s_b}\rho(s)P_N(s)ds
\end{equation}
where $[s_a,s_b]$ is any interval out of the total integration range. But
because the second term has now a polynomial weight, we can use global
duality to write
\begin{eqnarray}
\int_{s_a}^{s_b}\rho(s)P_N(s)ds
  &=&\frac1{2\pi i}\int_{s_a}^{s_b}{\sl Disc\,}\Pi(s)P_N(s)ds\ =\nonumber\\
  &=&-\frac1{2\pi i}\oint_{|s|=s_a}\Pi(-s)P_N(s)ds
  +\frac1{2\pi i}\oint_{|s|=s_b}\Pi(-s)P_N(s)ds.
\end{eqnarray}
Thus this part can be represented by a difference of two circle integrals
in the complex plane. On the other hand, the difference $H(s)-P_N(s)$
suppresses the contribution of the first part. Our method consists thus of
the following steps:
\begin{itemize}
\item replacing $\rho(s)$ in the first part of Eq.~(\ref{Groote:splitting})
by the value of the experimentally measured total cross section $R(s)$ (see
e.g.\ ref.~\cite{Groote:ref1})
\item replacing the circle integral contribution to flavours at their
threshold by zero
\item in all other cases inserting the QCD perturbative and non-perturbative
parts of $\Pi(-s)$ on the circle
\end{itemize}
These replacements can be seen as a concept within QCD sum rules. To obtain
the best efficience of our method, we have to restrict the polynomial
function by the following contraints:
\begin{itemize}
\item The method of least squares should be used to mimic the weight
\item However, the degree $N$ should not be higher than the order of the
highest perturbative resp.\ non-perturbative contribution increased by one
(this is a consequence of the Cauchy's theorem which is involved in the
analytical integration of the circle integrals)
\item Especially for the low energy region, the polynomial function should
vanish on the real axis to avoid instanton effects
\item In regions where resonances occur, the polynomial function should
fit the weight function to suppress those contributions which constitute the
highest uncertainty of the experimental data
\end{itemize}
As just mentioned, the integration on the circle can be done analytically
by using the Cauchy's theorem. But we have to keep in mind that the result
for $\Pi(-s)$ we use here depends logarithmically on the renormalization
scale $\mu$ and on the parameters of the theory that are renormalized at
the scale $\mu$. These are the strong coupling constant, the quark masses
and the condensates. As advocated in~\cite{Groote:refc}, we implement the 
renormalization group improvement for the moments of the electromagnetic 
correlator by performing the integrations over the circle with radius
$|s|=s_b$ with constant parameters, i.e.\ they are renormalized at a fixed
scale $\mu$. Subsequently these parameters are evolved from this scale to
$\mu^2=s_a$ using the four-loop $\beta$ function. In other words, we impose 
the renormalization group equation on the moments rather than on the
correlator itself. This procedure is not only technically simpler but also 
avoids possible inconsistencies inherent to the usual approach where one 
applies the renormalization group to the correlator, expands in powers of 
$\ln(s/\mu^2)$ and carries out the integration in the complex plane only 
at the end. In the present case the reference scale is given by 
$\Lambda_{\overline{\rm MS}}$.

\subsection*{Subdividing the integration interval}
As a first interval we select the range from the light flavour
production threshold $s_0=4m_\pi^2$ and the next threshold marked by the
mass of the $\psi$, $s_1=m_\psi^2\approx(3.1{\rm\,GeV})^2$. In this case we
set the inner circle integral to zero and obtain
\begin{equation}\label{Groote:sumrule}
\int_{s_0}^{s_1}R(s)H(s)ds=\int_{s_0}^{s_1}R^{\rm exp}(s)
  \left(H(s)-P_N(s)\right)ds+6\pi i\oint_{|s|=s_1}\Pi^{\rm QCD}(-s)P_N(s)ds.
\end{equation}
As mentioned above, we impose the constraints to the polynomial function
that it should vanish on the real axis at $s=s_1$ and should coincide with
the weight function at the $\rho$ resonance, i.e.\ for
$s=m_\rho^2\approx (1{\rm\,GeV})^2$.
\begin{figure}
\centerline{\epsfig{file=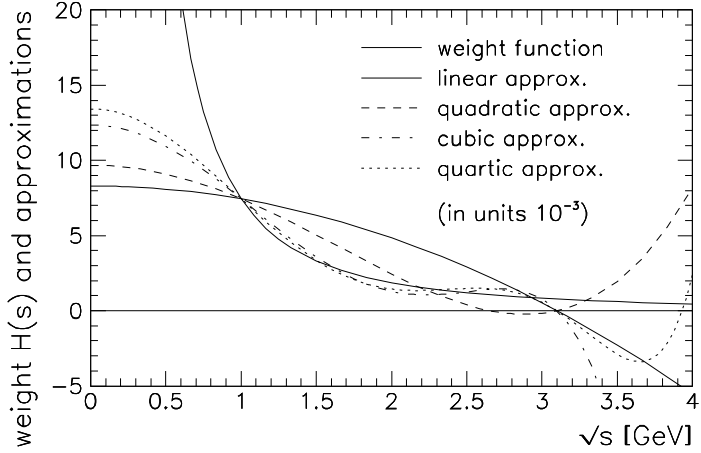}}
\vspace{10pt}
\caption{Weight function $H(s)$ and polynomial approximations $P_N(s)$ in
  the lowest energy interval $2m_\pi\le\sqrt s\le 3.1{\rm\,GeV}$. The least
  square fit was done in the interval $m_\rho\le\sqrt s\le 3.1{\rm\,GeV}$ with
  further constraints
  $H(s)=P_N(s)$ at $\sqrt s=1{\rm\,GeV}$ and $P_N(s)=0$ at
  $\sqrt s=3.1{\rm\,GeV}$. The quality of the polynomial approximations are
  shown up to $N=4$. We use the scaled variable $s/s_1$ for the polynomial
  approximation where $s_1$ is the upper radius such that $P_N(s/s_1)$ is
  dimensionless.}
\label{Groote:fig1}
\end{figure}
Fig.~\ref{Groote:fig1} shows polynomials of different order in comparison
with the weight function.
\begin{figure}
\centerline{\epsfig{file=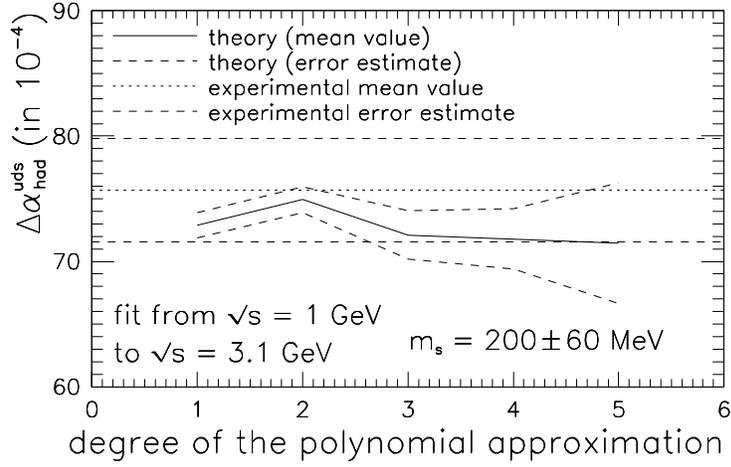}}
\vspace{10pt}
\caption{Comparison of the l.h.s.\ and r.h.s.\ of the sum rule given by 
  Eq.~(\ref{Groote:sumrule}) in the interval
  $0.28{\rm\,GeV}\le\sqrt s\le 3.1{\rm\,GeV}$. Dotted horizontal line: value
  of integrating the l.h.s.\ using experimental data including error 
  bars~\protect\cite{Groote:ref1}. The points give the values of the r.h.s.\
  integration for various orders $N$ of the polynomial approximation. 
  Straight line interpolations between the points are for illustration 
  only. The dashed lines indicate the error estimate of our calculation.}
\label{Groote:fig2}
\end{figure}
The results shown in Fig.~\ref{Groote:fig2} are compared with the result
obtained by using only the experimental data. For the up and down quarks we
only keep the mass independent part of the QCD contribution while for the
strange quark we include also the terms to order $O(m_s^2/q^2)$.

The second interval is limited by $s_1$ and the threshold marked by the
mass of the $\Upsilon$, $s_2=m_\Upsilon^2\approx(9.46{\rm\,GeV})^2$. For the
charm quark, we again set the inner circle integral to zero, but for the
lighter quarks we have to keep both. The perturbative series for the
charm quark is used up to it's known extend.

The third interval given between $s_2$ and $(40{\rm\,GeV})^2$ is again
subdivided into two pieces because of it's length. For the first of these
two intervals we choose $[(9.46{\rm\,GeV})^2,(30{\rm\,GeV})^2]$, for the
second $[(30{\rm\,GeV})^2,(40{\rm\,GeV})^2]$. Now the bottom quark is the
one for which the ``threshold rule'' (i.e.\ leaving out the inner circle)
applies. The remaining part of the integral starting from
$s_4=(40{\rm\,GeV})^2$ up to infinity is done using local duality, i.e.\
by inserting the function $R(s)$ obtained for perturbative QCD into the
second part of Eq.~(\ref{Groote:splitting}).

Our results are collected in Table~\ref{Groote:tab1}. To obtain these
results, we used the condensate values
\begin{equation}
\langle\frac{\alpha_s}\pi GG\rangle=(0.04\pm 0.04){\rm\,GeV}^4,\qquad
\alpha_s\langle\bar qq\rangle^2=(4\pm 4)\cdot 10^{-4}{\rm\,GeV}^6.
\end{equation}
For the errors coming from the uncertainty of the QCD scale we take
\begin{equation}
\Lambda_{\overline{\rm MS}}=380\pm 60{\rm\,MeV}
\end{equation}
The errors resulting from the uncertainty in the QCD scale in different 
energy intervals are clearly correlated and will have to be added linearly 
in the end. We also include the error of the strange quark mass in the 
light quark region which is taken as
\begin{equation}
\bar m_s(1{\rm\,GeV})=200\pm 60{\rm\,MeV}
\end{equation}
For the charm and bottom quark masses we use the values
\begin{equation}
\bar m_c(m_c)=1.4\pm 0.2{\rm\,GeV},\quad
\bar m_b(m_b)=4.8\pm 0.3{\rm\,GeV}.
\end{equation}
Summing up the contributions from the five flavours $u$, $d$, $s$, $c$, and 
$b$, our result for the hadronic contribution to the dispersion integral 
including the systematic error due to the dependence on 
$\Lambda_{\overline{\rm MS}}$ (column~5 in Table~\ref{Groote:tab1}) reads
\begin{equation}
\Delta\alpha_{\rm had}^{(5)}(M_Z)=(277.6\pm 4.1)\cdot 10^{-4}.
\end{equation}
In order to obtain the total result for $\alpha(M_Z)$, we have to add the 
lepton and top contributions. Since we have nothing new to add to the
calculation of these contributions we simply take the values from
ref.~\cite{Groote:ref5},
\begin{equation}
\Delta\alpha_{\rm had}^t(M_Z)=(-0.70\pm 0.05)\cdot 10^{-4},\qquad
\Delta\alpha_{\rm lep}(M_Z)\approx 314.97\cdot 10^{-4}.
\end{equation}
Writing 
$\Delta\alpha(M_Z)=\Delta\alpha_{\rm lep}(M_Z)+\Delta\alpha_{\rm had}(M_Z)$ 
our final result is ($\alpha(0)^{-1}=137.036$)
\begin{equation}
\alpha(M_Z)^{-1}=\alpha(0)^{-1}(1-\Delta\alpha(M_Z))=128.925\pm 0.056.
\end{equation}
\begin{table}
\caption{\label{Groote:tab1}Contributions of different energy intervals to
$\alpha_{\rm had}^{(5)}(M_Z)$. Second column: choice of neighbouring pairs
of the polynomial degree $N$. Third column: fraction of the contribution of
experimental data~\protect\cite{Groote:ref1}. Fourth column: contribution to
$\Delta\alpha_{\rm had}^{(5)}(M_Z)$ with all errors included except for the
systematic error due to the dependence on $\Lambda_{\overline{\rm MS}}$
which is separately listed in the fifth column.}
\begin{tabular}{rcrrr}
interval&values&data\qquad&contribution&error\\
for $\sqrt s$&of $N$&contribution
  &to $\Delta\alpha_{\rm had}^{(5)}(M_Z)$
    &due to $\Lambda_{\overline{\rm MS}}$\\[3pt]\hline
$[0.28{\rm\,GeV},3.1{\rm\,GeV}]$&$1,2$&$24\%$&$(73.9\pm 1.1)\cdot 10^{-4}$
  &$0.9\cdot 10^{-4}$\\
$[3.1{\rm\,GeV},9.46{\rm\,GeV}]$&$3,4$&$0.3\%$&$(69.5\pm 3.0)\cdot 10^{-4}$
  &$1.4\cdot 10^{-4}$\\
$[9.46{\rm\,GeV},30{\rm\,GeV}]$&$3,4$&$1.1\%$&$(71.6\pm 0.5)\cdot 10^{-4}$
  &$0.06\cdot 10^{-4}$\\
$[30{\rm\,GeV},40{\rm\,GeV}]$&$3,4$&$0.15\%$&$(19.93\pm 0.01)\cdot 10^{-4}$
  &$0.02\cdot 10^{-4}$\\
$\sqrt s>40{\rm\,GeV}$&&&$(42.67\pm 0.09)\cdot 10^{-4}$&\\\hline
total range&&&$(277.6\pm 3.2)\cdot 10^{-4}$&$1.67\cdot 10^{-4}$\\
\end{tabular}
\end{table}

\section*{Conclusion and outlook}
I have presented a method to obtain the running fine structure constant
$\alpha$ at the scale of the $Z_0$ mass with minimal input of experimental
data. This method is conservative in the meaning that its error is as free
from assumptions as possible. Our method is discussed and compared with
other methods (see e.g.\ ref.~\cite{Groote:refd} -- however, with our
preliminary results). In ref.~\cite{Groote:refe} Matthias Steinhauser says
that ``it is very impressive that the new analysis show very good agreement
both in their central values and their quoted errors.'' I nevertheless would
like to close this talk with the remark that all recent calculations of
$\alpha(M_Z)$ should not deter experimentalists from remeasuring the
$e^+e^-$ annihilation cross section more accurately in the low and
intermediate energy region, as such data are absolutely essential for a
precise value of $\alpha(M_Z)$, unbiased by theory.

\section*{Acknowledgements}
We would like to thank F.~Jegerlehner and R.~Harlander for providing us with
material in addition to refs.~\cite{Groote:ref1,Groote:ref7}, A.H.~Hoang for
correspondence, and G.~Quast for discussions on all experimental aspects of
this work and for continuing encouragement. S.G.\ gratefully acknowledges a
grant given by the Max Kade Foundation.

\end{document}